%% file: grb.tex
\documentclass[12pt]{article}
\usepackage{epsfig}

\input econfmacros.tex
\def\Title#1{\begin{center} {\Large {\bf #1} } \end{center}}

\begin{document}

\Title{MHD Jets, Flares, and Gamma Ray Bursts}
\bigskip\bigskip


\begin{raggedright}  

{\it Kazunari Shibata and Seiichiro Aoki \index{Shibata, K. and Aoki, S.}\\
Kwasan Observatory\\
Kyoto University\\
Yamashina, Kyoto 607-8471, JAPAN}
\bigskip\bigskip
\end{raggedright}

\section*{Abstract}

Recent numerical simulations of 
MHD jets from accretion disks are briefly reviewed with emphasis on
the scaling law for jet speed and the role 
of magnetic reconnection in relation to time variability
in accretion disks, jets, and flares. 
On the basis of these studies, 
possible interpretation is given on why statistical 
properties of peak intensity, peak interval, and 
peak duration of gamma ray bursts (log-normal distribution) are
different from those in solar flares
and black hole accretion disks (power-law distribution).
From these considerations, a new model, ``magnetized plasmoid model'',
is proposed for a central engine of gamma ray bursts.

\section{Introduction}

Recent development of astronomical observations has revealed that
our universe is full of enigmatic explosive phenomena, such as 
jets, bursts, and flares. Jets
ejected from active galactic nuclei (AGN) are probably among the biggest 
in size and the most energetic in total energy. 
Similar jets on much smaller scale
have been found in young stellar objects (YSO) as well as
in close binary system. 
These active objects 
show  vigorous time variability, often called bursts or flares, 
in optical, radio waves, X-rays, etc., 
in almost all electromagnetic spectrum. Though the central objects and jets
in AGNs, YSOs, and binary system are quite different in mass and size,
there are many similar properties in them, such as 
time variability, morphology of jets, and 
existence of accretion disks.

Gamma ray bursts (GRBs) were discovered nearly 30 years ago. Since then,
they have remained the most enigmatic object in our universe.
Recent rapid development of observations, however, is
uncovering a part of enigma of GRBs (e.g., Fishman and Meegan 1995, 
Piran 1999, Meszaros 2002);
(1) they occur in cosmological distance (i.e., most luminous in our universe,
$\sim 10^{51-53}$ erg/s, but different from AGNs), 
(2) there is evidence that they are emitted from
relativistic jets via synchrotron emission, and 
(3) some GRBs (long bursts) seem to be associated with supernovae. 
Some properties are similar to those of AGN jets (blazars), suggesting
common physics in blazars and GRBs. 

On the other hand, recent space observations of our Sun, the nearest star, 
have revealed that the solar corona is full of jets
and flares.  Though the total energy of the solar jets and flares
is much smaller than those in comic jets and flares,  the 
spectrum and time variability of electromagnetic waves emitted from
solar flares are quite similar to those of
cosmic flares, suggesting common physical origin. 
In the case of the solar flares,
it has been established that magnetic field is the source of
energy, so that the knowledge of solar jets
and flares will be useful for understanding the role of magnetic field
in cosmic jets and flares in distant stars and galaxies. 
 
In this article, we first briefly
review recent understanding of magnetically driven jets from 
accretion disks (as a model of AGN jets). We then discuss flares and 
magnetic reconnection associated with production of jets, especially 
in the case of protostars.
Finally, we discuss similarity and differences in time variability 
of GRBs, black hole accretion disks, and solar flares, and on the basis of
these observations, we propose a new 
model for central engine of GRBs.

\section{MHD Jets}

Magnetically driven jets from accretion disks (Fig. 1) were 
first proposed to explain jets from active galactic nuclei 
(Blandford 1976, Lovelace 1976,  Blandford and Payne 1982). 
After the discovery of CO molecular bipolar flows in 
star forming regions (Snell et al. 1980), 
the MHD jet model started to be applied to bipolar flows and jets 
from young stars (Uchida and Shibata 1985, Pudritz and Norman 1986).
The first time-dependent numerical simulation of MHD jets from
accretion disks were carried out by Uchida and Shibata (1985)
and Shibata and Uchida (1986).

The MHD jet model (e.g., Tajima and Shibata 1997, Ferrari 1998, for a review)
has the following merits:
(1) the magnetic force not only {\it accelerate} 
plasmas from disk surface to
form bipolar jets  but also {\it extract angular momentum}
 from accretion disks, 
enabling efficient accretion of plasma
onto central objects (stars or black holes),
(2) the magnetic force due to toroidal fields 
{\it collimate} jets by pinching effect.

\begin{figure}[htb]
\begin{center}
\caption{Typical example of 2.5D MHD numerical simulations
MHD jets from a thick disk (Kudoh et al. 2002).
Note that the disk becomes turbulent because of 
magneto-rotational instability.
}
\label{fig:kudoh_thick}
\end{center}
\end{figure}

Shibata and Uchida (1986) and subsequent
many 2.5D MHD numerical simulations 
(Stone and Norman 1994,
Matsumoto et al. 1996, Hirose et al. 1997, Ouyed and Pudritz 1997, 
Kudoh et al. 1998, 2002, Kuwabara et al. 1999, Ustyugova et al.
1999, Kato et al. 2002) and 
3D MHD simulations (Matsumoto and Shibata 1997, 
Ouyed et al. 2003) have revealed that
{\it the velocity of MHD jets is of order of Keplerian velocity at the
footpoint of MHD jets}.  Kudoh et al. (1998) and Kato et al. (2002)
have shown that the jet velocity
has  a weak dependence on $B$ (Fig. 2);
$$ V_{jet} \propto {E_{mg}}^{1/6} \propto  {B_p}^{1/3}.  \eqno(1)       $$
Here,$E_{mg} = (V_A/V_k)^2 \propto {B_p}^2$, and $B_p$ is the initial poloidal
magnetic field strength.

This is consistent with the results of one-dimensional steady jet theory
(Kudoh and Shibata 1995, 1997), and also can be 
derived semi-analytically using the Michel (1969)'s relation;
$$ V_{\infty} \simeq \Bigl( {\Omega^2 B_p^2 r^4 \over \dot M} \Bigr)^{1/3}.
                                     \eqno(2)  $$
Here, $\Omega$ is the angular speed of the disk at the footpoint of the jet, 
$r$ is the radial distance from the central object, and $\dot M$ is the
mass flux of the jet.

\begin{figure}[htb]
\begin{center}
\caption{(a) Maximum velocity of jets in unit of Keplerian
velocity at the footpoint of jets.
(b) Mass flux of jets. (c) Mass accretion rate.
(d) Ratio of mass flux of jets to mass accretion rate.
All data are based on 2.5D nonsteady MHD simulations
in the case of thick disks  (Kudoh et al. 1998)
}
\label{fig:kudoh_max_velocity}
\end{center}
\end{figure}

It is very important to note that the terminal velocity strongly depends
on the mass flux $\dot M$.  
In our problem, the mass flux is given by
$$ \dot M \simeq 4 \pi \rho_{slow} V_{slow} r^2 
         \simeq 4 \pi \rho_{slow} C_s {B_p \over B} r^2,   \eqno(3) $$
where $\rho_{slow}$ is the mass density 
at the {\it slow magnetosonic point},
$V_{slow}$ is the slow magnetosonic speed, $C_s$ is the sound speed,
and $B = (B_p^2 + B_{\varphi}^2)^{1/2}$. 
In a cold disk, the slow magnetosonic point corresponds to the local
maximum of the effective gravitational (Blandford-Payne) potential, 
and is located near the disk plane ($\rho_{slow}/\rho_0 
\sim 0.1$ in our case). 
 For stronger fields ($E_{mg} > 10^{-2}$),
the magnetic field lines become straight $B \sim B_p \gg B_{\varphi}$ so that 
the mass flux does not depend on $B_p$, but for weaker fields
($E_{mg} < 10^{-2}$),
field lines are highly twisted and the azimuthal component becomes
dominant $B \sim B_{\varphi} \gg B_p$ near the disk so that the mass flux
is in proportion to $B_p$ (Kudoh and Shibata 1995, 1997).
Considering these effects, the terminal speed of the MHD jet becomes
$$ {V_{\infty} \over V_k} \simeq 
\Bigl({\rho_{slow} \over \rho_0 }\Bigr)^{-1/6}
E_{th}^{-1/6}  E_{mg}^{1/6},    \eqno(4)    $$
where 
$ E_{th} = (C_s/V_k)^2$ = (thermal energy)/ 
                   (gravitational energy), 
$ E_{mg} = (V_A/V_k)^2$ = (magnetic energy)/ 
                   (gravitational energy).
Since $\rho_{slow}/\rho_0 \sim 0.1$ and $E_{th}/E_{mg} \simeq \beta
=$ gas pressure / magnetic pressure $\sim 1 - 10$ in the disk,
 we find that 
{\it the terminal speed of the jet is comparable to
the Keplerian velocity} for wide range of
poloidal magnetic field strength, 
and $V_{\infty} \propto B_p^{1/3}$. This explains the results (eq. 1) 
of the previous 2.5D MHD numerical simulations  very well.

We should emphasize again that {\it 
even if the magnetic field strenth  is
very weak in accretion disks, the jet velocity is roughly comparable to
Keplerian speed ($V_{jet} \sim 0.1 - 1.0 V_k$ for $E_{mg} \sim
10^{-8} - 10^{-2}$)}. The physical reason is that
magnetic field lines are highly twisted by the differential rotation of
the disk until the local magnetic energy density ($B_{\varphi}^2/8\pi$)
 becomes comparable to
the rotational energy ($\rho V_k^2/2 \sim \rho GM/r $ 
gravitational energy) at the surface of
the disk. Since the kinetic
energy of the jet ($\rho V_{jet}^2/2$) comes from the magnetic energy, 
it eventually becomes comparable
to the gravitational energy ($\rho V_k^2/2$)
at the disk surface (i.e., at the 
slow magnetosonic point). This process is
similar to the magneto-rotational instability (Balbus and Hawley 1991) 
in the sense that the magnetic effect  
becomes eventually important even if the initial magnetic field is very weak.

Recently, Koide et al. (1998, 2000, 2002) have succeeded to
extend these newtonian MHD simulations of jets
to general relativistic MHD versions.
They have shown that the accretion (and the jet ejection) become more
violent near black hole than in the newtonian case.
They confirmed the extraction of rotational energy from a Kerr hole
by the effect of magnetic field. 
The maximum Lorentz factor of jets in these simulations is still
of order of 2, much smaller than those observed for AGN jets 
(Lorentz factor 10-100) and GRBs (Lorentz factor 100-1000).
We do not know yet whether this result (small maximum Lorentz factor) is
simply a result of numerical limitations, or a result of physics. 

It is important to note that the jet ejection has never reached steady
state, even if the velocity and mass flux of the jet is well explained
by steady theory (Kudoh et al. 1998).  Recently, Sato et al. (2003) have
succeeded to run the MHD simulation of jets including accretion disk 
self-consistently for many
orbital periods (up to 15-20 orbits), and revealed that the jet ejection
is intermittent and often associated with transient accretion events.
They found that accretion disk is fully turbulent (due to
magneto-rotational instability), and full of reconnection event.
Often the jets are ejected in association with such reconnection events.
The time variability of mass accretion rate in a simulated accretion disk 
show power law in the power density spectrum
as first noted by Kawaguchi et al. (2000).

\section{Flares : Magnetic Reconnection}

Recent space solar observations such as Yohkoh, SOHO, TRACE
have revealed that {\it solar corona is much more dynamic than had
been thought, and is full of flares, microflares, nanoflares, jets, and 
various mass ejections}. Among them, the largest mass ejectins are called
coronal mass ejections (CMEs). It has also been revealed that 
 that the reconnection
plays essential role not only in large scale flares and CMEs, but also
small scale flares and jets, leading to unified model (e.g., Shibata 1999).

\begin{figure}[htb]
\begin{center}
\caption{
A magnetic reconnection model of protostellar flares 
(Hayashi et al. 1996). It is assumed that at 
$t =0$ a stellar dipole magnetic field penetrates an accretion disk.
The parameters in the initial disk (at $r=1$) 
are $E_{th} = 2 \times 10^{-3},
E_{mg} = 2 \times 10^{-4}$. 
The color shows the temperature, and solid curves denote magnetic
field lines. The arrows depict velocity vectors in r-z plane.
}
\label{fig:hayashi}
\end{center}
\end{figure}

Hayashi, Shibata, Matsumoto (1996) presented a magnetic reconnection
model of protostellar flares, by performing 2.5D time dependent MHD
nummerical simulations of interaction between an accretion disk
and stellar dipole magnetic field.
Figure 3 shows one of their simulation results.
They assumed that an accretion disk
is penetrated by stellar dipole field at t = 0, and examined
the subsequent evolution of the interaction
between a rotating disk and a stellar dipole field.
The initial process occurring near the disk is basically
the same as those in the nonsteady MHD jet model (e.g.,
Shibata and Uchida 1986). The magnetic field is twisted by the rotating
disk, and the J x B force associated with the twist 
accelerates the plasma in the surface layer of the disk to form an MHD
jet in bipolar directions. In this case,  the magnetic
twist is accumulated in a closed loop, 
increasing the magnetic pressure of the loop, which eventually
leads to the ejection of the magnetic loop after about 
one orbit. 
After the ejection of the loop, a current sheet
is created inside the loop, leading to fast reconnection there.
This process is similar to that occurring in 
solar coronal mass ejections, and basic reconnection mechanism
is the same as in solar flares (e.g., Tsuneta et al. 1992, 
Shibata 1999, Yokoyama and Shibata 2001). 
The reconnection releases huge
amount of magnetic energy of order of  $10^{36}$ erg
(about $10^4$ times more energetic than solar flares)
stored in a sheared loop with 
a size of $L \sim 10^{11}$ cm. 

The temperature of super hot
plasma created by reconnection amounts to 
$$ T \sim 10^8 \Bigl({B \over 100 {\rm G}} \Bigr)^{6/7}
     \Bigl({n_0 \over 10^9 {\rm cm}^{-3}} \Bigr)^{-1/7}
     \Bigl({L \over 10^{11.5} {\rm cm}} \Bigr)^{2/7}
       \ \ \ {\rm K},  \eqno(5) $$
which is based on the balance between reconnection heating
and conduction cooling 
(Yokoyama and Shibata 2001, Shibata and Yokoyama 2002). 
Here $n_0$ is the pre-flare coronal density. 
These results explain
characteristics of protostellar flares observed by
ASCA and ROSAT (e.g., Koyama et al. 1996, Shibata and Yokoyama 2002).

\section{Gamma Ray Bursts}

It is well known that the gamma ray
burst light curves are very similar to those of solar flare gamma rays 
and hard X-rays.
Hence, it was often considered that there may be some physical 
similarity between gamma ray bursts and solar flares.
Recently, it has in fact been revealed that 
there is evidence that gamma ray bursts are emitted from 
collimated jets, which
might be similar to relativistic jets from active galactic nuclei. 
If this is so,  MHD jet models developed for AGN jets can be 
applied to
gamma ray bursts.  There is also evidence that some neutron stars have 
very strong magnetic fields up to $10^{15}$ G, 
called magnetars (Duncan and Thompson 1992, Kouveliotou et al. 1999). 
Hence, the GRB model including 
strong magnetic field has been  proposed (e.g., Kluzniak 
and Ruderman 1998). 
On the other hand, recent theory of accretion disks
showed that magnetic field is essential for generating viscosity
through the occurrence of magneto-rotational instability (Balbus and
Hawley 1991). 
Altogether, the role of magnetic field in gamma ray bursts is 
considered to be more and more important
than had been thought.  

Here we shall discuss analogy between solar flares/coronal mass 
ejections (CME) and
gamma ray bursts, emphasizing the basic MHD physics of solar flares/CMEs.

\begin{figure}[htb]
\begin{center}
\caption{Ligh curves of (a) GRBs
(http://www.batse.msfc.nasa.gov/batse/grb/lightcurve),  
(d) black hole accretion disks (BH-ADs, Miyamoto et al.
1992), 
(g) solar flares (Zirin et al. 1971), 
power density spectra of 
(b) GRBs (Belobodorov et al. 2002), 
(d) BH-ADs (Miyamoto et al. 1992), 
(h)solar flares (Ueno et al. 1997), 
and histograms of time intervals 
of (c) GRBs (Nakar and Piran 2002), 
(f) BH-ADs (Negoro et al. 1995), 
and (i) solar flares (Wheatland 2000).  }
\label{fig:grb_power}
\end{center}
\end{figure}

As we discussed above, the time variability of 
gamma ray burst light curve is similar to
those of solar flares.  The power spectrum analysis of
time variability of gamma ray bursts (Beloborodov et al. 2000) and 
solar X-ray emission (Ueno et al. 1997) show
that both show  power-law disribution with
index $\alpha \sim 1.5-1.8$. 
Interestingly, the X-ray light curve of Cygnus X-1
(black hole candidate) also show similar time variability with power law
spectrum (Miyamoto et al. 1992, Negoro 1992).

\begin{figure}[htb]
\begin{center}
\caption{(a) Number of CMEs associated with solar flares vs. 
peak X-ray intensity of these flares, 
(b) number of CMEs associated with solar flares vs. 
peak interval of these flares (from Aoki et al. 2003).}
\label{fig:cme}
\end{center}
\end{figure}

However, there is a fundamental difference in statistical properties 
between gamma ray bursts and solar flares.
That is, many physical quantities in solar flares, 
such as duration, peak intensity,
peak interval, show power-law frequency distributions, while these quantities
in gamma ray bursts do not show power-law  distribution but show
log-normal distribution (Li and Fenimore 1996; see Fig. 4).  
In other words, there is no characteristic 
time and intensity in solar flares, whereas there is characteristic 
time and intensity in gamma ray bursts.
What does this mean ?   In spite of apparent similarity between solar flares
and gamma ray bursts, does this suggest that basic physics of gamma ray bursts
is different from that of solar flares ? 

In relation to this, Negoro and Mineshige (2002) recently found interesting
fact on statistical properties of X-ray shots (sub-bursts) emitted from 
black hole accretion disks: If only large shots are picked up from the time 
variability of X-ray emission of accretion disk, 
the distribution of shot peak become log-normal.
They argued that the X-ray shots are occurring at the surface of accresion 
disks, and hence may be similar 
to solar flares and do not have characteristic time scale
(Ueno et al. 1997, Kawaguchi et al. 2000).
On the other hand, according to internal shock model of gamma ray bursts
(e.g., Meszaros 2002),
the burst emissions are from internal shocks in the jet ejected
from central engines.  If the jets are ejected in association with
X-ray shots, only large shots can eject enough
mass to generate internal shock. This explains log-normal distribution 
in the duration and peak intensity of GRB.

Similar discussion may be applied to solar flares.  
In the case of solar flares,
there is a tendency that larger flares produce larger mass ejections, called
coronal mass ejections (CMEs).  
It is difficult for small flares to eject much mass
from solar surface, because they do not have enough energy to escape from
magnetically confined solar active regions.  
Aoki et al. (2003) examined this property by
using actual data. Figure 5 shows number of CMEs associated with solar flares
versus peak X-ray flux of those flares, indicating that log-normal distribution
roughly holds.  The number of CMEs versus CME speeds also show 
similar log-normal distribution. 

Hence we propose a new model for gamma ray bursts (see Figure 6), which
is basically in the same line of thought as that of 
Negoro and Mineshige (2002): 
magnetic reconnection (flares) occurs everywhere in the 
surface of accretion disks,
whose spatial distribution is fractal, and time variability 
is power-law, both of 
which are quite similar to those of solar flares.  Only ejecta from 
energetically and
spatially large reconnection events (flares or shots) can escape from 
magnetosphere
of accretion disks to form jets. In fact, nonsteady MHD simulations of 
magnetically
driven jets from accretion diks (Kudoh et al. 2002, Kuwabara et al. 2002, 
Sato et al.
2003) show that the ejection of jets is highly time variable, intermittent, and
base of jets is full of reconnection events; when large reconnection 
events occur,
large plasmoid ejection occurs, like the protostellar flare model 
(Hayashi et al. 1996, Miller and Stone 1997, Goodson et al. 1999).
Jets have non-uniform density distribution, consisting of
intermittently ejected plasmoid (confined in 
magnetic island or helical field in 3D space) like
solar coronal mass ejections.  It is natural that these intermittently ejected
magnetized plasmoid produce lots of internal shocks, which are site of gamma 
ray emission as modeled in internal shock model of gamma ray bursts.

\begin{figure}[htb]
\begin{center}
\caption{Schematic illustration of ``magnetized plasmoid model''
(or flare/CME model) of gamma ray bursts
(Aoki et al. 2003).}
\label{fig:grb_model}
\end{center}
\end{figure}

There is another merit in this ``magnetized plasmoid model'' 
(or ``flare/CME model''). 
Ejected plasmoid has a structure similar to spheromak.
It is well known that spheromak is unstable to tilting instability
(e.g., Hayashi and Sato 1984),
so that ejected spheromak-like plasmoids would have various angles to the
direction of the ejection as they propagate. 
Hence they collide with each other, and there is a high possibility
that tangent magnetic field has an opposite component, thus leading to
magnetic reconnection. This means that even if the initial
energy conversion from magnetic energy to plasma energy (kinetic and
internal) is not efficient, it is possible that all magnetic energy
(Poynting flux) contained in magnetized plasmoids would eventually be
converted to plasma energy through magnetic reconnection.
It should also be noted that magnetic reconnection generate lots of
MHD shocks: Petschek slow shocks are formed just sides of reconnection
jets, and fast shocks are created when reconnection jets collide with
ambient medium (e.g., Yokoyama and Shibata 2001). 
These situations, containing lots of shocks as well
as X-type and O-type neutral points, are similar to fractal MHD
turbulence, and very suitable for high energy particle acceleration.
Such fractal structure or fractal reconnection (Shibata and Tanuma 2001,
Tanuma et al. 2001) may be the origin of power-law time variability spectrum
of solar flares, black hole accretion disks, and GRBs.

\bigskip

We are grateful to T. Kudoh, S. X. Kato, K. Sato,  
S. Mineshige, H. Negoro, T. Murakami, and T. Ishii 
for various help and useful discussions.

\def\Discussion{
\setlength{\parskip}{0.3cm}\setlength{\parindent}{0.0cm}
     \bigskip\bigskip      {\Large {\bf Discussion}} \bigskip}
\def\speaker#1{{\bf #1:}\ }
\def\endDiscussion{}

\Discussion

\speaker{J. Rhoads}
How do you  generalize the relation
$V_{jet} = v_{escape} = v_{Kepler}$
to the highly relativistic case ?  Here $V_{jet} = v_{escape} = c$ is obvious,
but it is not clear
if $\Gamma_{jet} = \Gamma_{escape}$. For the Newtonian case, I assume that
$V_{escape}$ is
measured in the region where the jet is launched. Trying to apply the
same idea to the $\Gamma= 100$ flow in a GRB would imply the jet is
launched barely outside the event horizon.  This seems improbable. 
How do we explain this ?

\speaker{K. Shibata}
You are right. In the Newtonian case, our theory explains observed
speeds of various jets very well, such as 
protostellar jets, SS433 jets, jets from cataclysmic variables, and so on.
However, in the relativistic case, we have not yet succeeded
to explain observed large Lorentz factor of AGN jets and gamma ray bursts.

\speaker{M. Lyutikov}
In case of a rotating BH the flow is generated on field lines that do not
cross BH horizon, but thread the ergosphere.

\speaker{K. Shibata}
Koide et al. (2002) successfully simulated such case.

\speaker{A. Beloborodov}
Keplerian timescale is very short (~ms) to explain the main variability in
GRBs (0.1-10 s). Should the accretion timescale appear as a characteristic
time in the model variability spectra ?

\speaker{K. Shibata}
MHD numerical simulations of accretion disks and jets revealed that
the time variability of accretion rate show temporal $1/f^{\alpha}$
fluctuation (Kawaguchi et al. 2000, Sato et al. 2003), i.e., there are
continuous distribution on time scales, which includes the one much longer
than Keplerian time scale. 
The longest time scale of variability is probably related to 
duration of main accretion and/or dynamo 
in our magnetized accretion disks. 

\speaker{S. Woosley}
Getting 10\% of the accreted mass into the jet is too much in the context
of GRBs.
$\Gamma$ of 2 may be OK initially but very important is the energy loading
of that matter.
Do you follow energy generation or are your jets all "cold" ? 
Do you see the energy loading might be increased ?

\speaker{K. Shibata}
The jet with 10\% of the accreted mass is a cold, dense jet directly ejected
from cold part of the disk. We found another component of a jet ejected 
from hot corona with less material.
Since our mechanism can produce enough poynting flux to explain total
energy of GRBs, we suggest magnetic reconnection in the corona and jet 
might explain the energy loading.
Such simulations of jets including reconnection in general relativistic
regime should be done in future.

\speaker{A. Hujeirat}
The Blandford \& Payne 1982 and my calculations predict the existence of a
super-Keplerian transition layer between the disk and the corona. Why do
your calculation don't predict such a layer ?

\speaker{K. Shibata}
Our calculations also show such super-Keplerian layer just above the disk
(see e.g., Shibata and Uchida 1986, Kudoh and Shibata 1997).

\speaker{A. Brandenburg}
Why  does the magnetic field not penetrate the black hole horizon?  This
seems
to be quite a general phenomena in relativistic electrodynamics.  What is
the force preventing this ?

\speaker{K. Shibata}
In our simulations, we adopt the coordinate system such that the time
proceeds slowly near the event horizon, so that accreting 
plasmas and magnetic field lines have not yet penetrated into 
the event horizon in our results.

\speaker{C. Fendt}
Your statement about the highly intermittant jet formation was derived
from ideal MHD simulations of the disk. What happens if you include
turbulence or diffusion ?  Will it stabilize the jet formation?

\speaker{K. Shibata}
Kuwabara et al. (2002) studied the effect of resistivity on jet formation. 
According to them, if the resistivity is large 
($R_m = rV_k/\eta < 100$), the jet
formation is suppressed. In the interemediate regime 
($100 < R_m < 1000$), the quasi-steady jet 
is formed. In the case of weak resistivity ($R_m > 1000$), 
the results show formation of
jets with high time variability and intermittency.

\endDiscussion
 
\end{document}

%% file: econfmacros.tex
\def\beq{\begin{equation}}
\def\eeq#1{\label{#1}\end{equation}}
\def\eeqn{\end{equation}}

\def\beqa{\begin{eqnarray}}
\def\eeqa#1{\label{#1}\end{eqnarray}}
\def\eeqan{\end{eqnarray}}

\let\bar=\overbar

\def\Dslash{\not{\hbox{\kern-4pt $D$}}}
\def\dslash{\not{\hbox{\kern-2pt $\del$}}}

\def\msb{{\bar{\ssstyle M \kern -1pt S}}}




%% file: grb.bbl
\begin{thebibliography}{99}


\bibitem{Aoki}
Aoki, S., Yashiro, S.\ and Shibata, K., to be submitted (2003)

\bibitem{Balbus}
Balbus, S. A. \  and Hawley, J. F., ApJ 376, 214 (1991)

\bibitem{Beloborodov}
Beloborodov, A. M., Stern, B. E.\ and Svensson, R., 
ApJ 535, 158 (2000)


\bibitem{Blandford76}
Blandford, R. D., MNRAS 176, 465 (1976)

\bibitem{Blandford-Payne}
Blandford, R. D. \ and Payne, D. G., MNRAS 199, 883 (1982)


\bibitem{Duncan}
Duncan, R. C.\ and Thompson, C., ApJ 392, L9 (1992)

\bibitem{Ferrari}
Ferrari, A., ARAA 36, 539 (1998)

\bibitem{Fishman}
Fishman, G. J.\ and Meegan, C. A., ARAA 33, 415 (1995)


\bibitem{Hayashi}
Hayashi, M. R., Shibata, K. \ and Matsumoto, R.,
  ApJ 468, L37 (1996)


\bibitem{tilting}
Hayashi, T.\ and Sato, T., Phys. Fluids 27, 778 (1984)

\bibitem{Hirose}
Hirose, S., Uchida, Y., Shibata, K.\ and Matsumoto, R.,
   PASJ 49, 193 (1997)


\bibitem{Kato}
Kato, S. X., Kudoh, T. \ and Shibata, K., ApJ 565, 1035 (2002)

\bibitem{Li}
Li, H.\ and Fenimore, E. E., ApJ 469, L115 (1996)

\bibitem{Kawaguchi}
Kawaguchi, T., et al., 
PASJ 52, L1 (2000)

\bibitem{Kluzniak}
Kluzniak, W.\ and Ruderman, M., ApJ 408, 179 (1998)

\bibitem{Koide-1}
Koide, S., Shibata, K.\ Kudoh, T., ApJ 495, L63 (1998)

\bibitem{Koide-2}
Koide, S., Meier, D., Kudoh, T.\ and Shibata, K., 
ApJ 536, 668 (2000)

\bibitem{Koide-3}
Koide, S., Shibata, K., Kudoh, T.\ and Meier, D. L., 
Science  295, 1688 (2002)

\bibitem{Kouvelitou}
Kouveliotou, C., et al., ApJ 510, L115 (1999)


\bibitem{Koyama}
Koyama, K., et al., PASJ 48, L87 (1996)


\bibitem{Kudoh95}
Kudoh, T.\ and Shibata, K., ApJ 452, L41 (1995)

\bibitem{Kudoh97a}
Kudoh, T.\ and Shibata, K., ApJ 474, 362 (1997)

\bibitem{Kudoh98}
Kudoh, T., Matsumoto, R.\ and Shibata, K., ApJ 508, 186 (1998)

\bibitem{Kudoh02}
Kudoh, T., Matsumoto, R.\ and Shibata, K., PASJ 54, 121 (2002)


\bibitem{Kuwabara00}
Kuwabara, T., et al.,  
PASJ 52, 1109 (2000)


\bibitem{Lovelace76}
Lovelace, R. V. E., Nature 262, 649 (1976)


\bibitem{Matsumoto96}
Matsumoto, R., et al., 
ApJ 461, 115 (1996)


\bibitem{Matsumoto97}
Matsumoto, R.\ and Shibata, K.,
in Proc. {\it Accretion Phenomena and Related Outflows}, 
IAU Colloq. No. 163, PASP
Conf. Ser. Vol. 121, Wickramasinghe, D. T. et al. (eds.), p. 443 (1997)

\bibitem{Meszaros}
Meszaros, P., ARAA 40, 137 (2002)


\bibitem{Michel}
Michel, F. C.,  ApJ 158, 727 (1969)


\bibitem{Miyamoto}
Miyamoto, S., et al., ApJ 391, L21 (1992)

\bibitem{Nakar}
Nakar, E.\ and Piran, T., MNRAS 331, 40 (2002)


\bibitem{Negoro}
Negoro, H., Ph. D. Thesis, Osaka University (1992)

\bibitem{Negoro}
Negoro, H., et al., ApJ 452, L49 (1995)

\bibitem{Negoro}
Negoro, H.\ and Mineshige, S., PASJ 54, L69 (2002)


\bibitem{Ouyed}
Ouyed, R.\ and Pudritz, R. E., ApJ 482, 712 (1997)

\bibitem{Ouyed3}
Ouyed, R., Clark, D. A.\ and Pudritz, R, E., ApJ 582, 292 (2003)



\bibitem{Pudritz}
Pudritz, R. E.\ and Norman, C., ApJ 301, 571 (1986)

\bibitem{Pudritz}
Piran, T., Physics Reports 314, 575 (1999)


\bibitem{sato}
Sato, K., et al., 
to be submitted (2003)

\bibitem{Shibata-Uchida86}
Shibata, K.\ and Uchida, Y., PASJ 38, 631 (1986)


\bibitem{Shibata99}
Shibata, K., Ap. Sp. Sci. 264, 129 (1999)

\bibitem{Shibata-Tanuma}
Shibata, K. \ and Tanuma, S., Earth, Planets, and Space 53, 473 (2001)

\bibitem{Shibata-Yokoyama02}
Shibata, K. \ and Yokoyama, T., ApJ 577, 422 (2002) 


\bibitem{Snell}
Snell, R. L., Loren, R. B.\ and Plambeck, R. L., 
ApJ 239, L17 (1980)


\bibitem{Stone94}
Stone, J. M.\ and Norman, M., ApJ 433, 746 (1994)

\bibitem{Tajima-Shibata}
Tajima, T.\ and Shibata, K., 
{\it Plasma Astrophysics},  Addison Wesley (1997)

\bibitem{Tanuma}
Tanuma, S., et al., ApJ 551, 312 (2001)


\bibitem{Tsuneta92}
Tsuneta, S. et al., PASJ 44, L63 (1992)

\bibitem{Uchida-Shibata85}
Uchida, Y.\ and Shibata, K., PASJ 37, 515 (1985)

\bibitem{Ueno97}
Ueno, S., et al., ApJ 484, 920 (1997)


\bibitem{Ustyugova99}
Ustyugova, G. V., et al., 
ApJ 516, 221 (1999)

\bibitem{Yokoyama-Shibata01}
Yokoyama, T.\ and Shibata, K., ApJ 549, 1160 (2001)


\bibitem{Wheatland}
Wheatland, M. S., ApJ 536, L109 (2000)
 
\bibitem{Zirin}
Zirin, H., Pruss, G.\ and Vorpahl, J., Solar Phys. 19, 463 (1971)


\vspace{-1mm} 


\end{thebibliography}
